\def\rfr#1{eq. (\ref{#1})}

\def\dert#1#2{\frac{{{d}}{#1}}{{{d}}{#2}}}              

\def\bar{\begin{eqnarray}}
\def\ear{\end{eqnarray}}
\def\bb{\bibitem}
\def\eqi{\begin{equation}}
\def\eqf{\end{equation}}
\def\eqia{\begin{eqnarray}}
\def\eqfa{\end{eqnarray}}
\def\rp#1#2{{#1\over#2}}

\def\lb#1{\label{#1}}

\def\oc2{$\mathcal{O}(c^{-2})$}

\documentclass[11pt]{article}
\usepackage{amsmath,amsthm,amscd,amssymb}
\usepackage{latexsym}
\usepackage{graphicx,epsfig}

\begin{document}

\noindent{\bf \LARGE{On a recently proposed scalar-tensor-vector
metric extension of general relativity to explain the Pioneer
anomaly}}
\\
\\
\\
{Lorenzo Iorio}\\
{\it Viale Unit$\grave{a}$ di Italia 68, 70125\\Bari, Italy
\\tel./fax 0039 080 5443144
\\e-mail: lorenzo.iorio@libero.it}

\begin{abstract}
Recently, Brownstein and Moffat proposed a gravitational mechanism
to explain the Pioneer anomaly based on their scalar-tensor-vector
(STVG) metric theory of gravity. In this paper we show that their
model, fitted to the presently available data for the anomalous
Pioneer 10/11 acceleration, is in contrast with the latest
determinations of the extra-precessions of perihelia for Jupiter,
Saturn and Uranus.
\end{abstract}



\section{Introduction}
The so-called Pioneer anomaly (Anderson et al. 1998; 2002)
consists of an unexpected, almost constant and uniform
acceleration  directed towards the Sun \eqi A_{\rm Pio}=(8.74\pm
1.33)\times 10^{-10} \ {\rm m\ s}^{-2}\lb{pioa}\eqf detected in
the data of both the spacecraft Pioneer 10 (launched in March
1972) and Pioneer 11 (launched in April 1973) after they passed
the threshold of 20 Astronomical Units (AU; 1 AU is slightly less
than the average Earth-Sun distance and amounts to about 150
millions kilometers), although it might also have started to occur
after 10 AU only, according to a recent analysis of the Pioneer 11
data (Nieto and Anderson 2005). Latest communications with the
Pioneer spacecraft, confirming the persistence of such an
anomalous feature, occurred when they reached 40 AU (Pioneer 11)
and 70 AU (Pioneer 10).

This effect has recently attracted considerable attention because
of the possibility that it is a signal of some failure in the
currently accepted Newton-Einstein laws of gravitation (for a
review of some of the proposed mechanisms of gravitational origin
se e.g. (Dittus et al. 2005)); indeed, at present no convincing
explanations of it in terms of some non-gravitational effects
peculiar to the spacecraft themselves have yet been found.

In this paper we focus on one of the most recent attempts to find
a gravitational explanation  for the anomalous behavior of Pioneer
10/11 and discuss its validity by performing a clean and
unambiguous independent test by analyzing the observationally
determined extra-rates of perihelia of Jupiter, Saturn and Uranus.

\section{The predicted orbital effects}
In order to explain the Pioneer anomaly, Brownstein and Moffat
(2006), in the context of the STVG metric theory of gravitation by
Moffat (2006), consider a variation with distance of the Newtonian
gravitational constant $G(r)$ and propose the following radial
extra-acceleration affecting the motion of a test particle in the
weak field of a central mass $M$ \eqi A_{\rm BM}=-\rp{G_0
M\zeta(r)}{r^2}\left\{1-\exp\left[-\rp{r}{\lambda(r)}\right]\left[1+\rp{r}{\lambda(r)}\right]\right\}.\lb{accel}\eqf
Here $G_0$ is the `bare' value of the Newtonian gravitational
constant. Lacking at present a solution for $\zeta(r)$ and
$\lambda(r)$, the following parameterization is introduced for
them\footnote{In the notation of Brownstein and Moffat (2006)
$\zeta(r)$ and $d$ are $\alpha(r)$ and $\overline{r}$,
respectively. Note that there is an error in eq. (12) for
$\lambda(r)$, p. 3430 of (Brownstein and Moffat 2006): a $-$ sign
is lacking in front of $b$. Instead, eq. (27) of (Moffat 2007)
gives the correct expression.}
\begin{equation}\left\{\begin{array}{lll}\zeta(r)=\zeta_{\infty}\left[1-\exp\left(-\rp{r}{d}\right)\right]^{\rp{b}{2}},\\\\
\lambda(r)=\lambda_{\infty}\left[1-\exp\left(-\rp{r}{d}\right)\right]^{-b}.\lb{al}\end{array}\right.\end{equation}
In \rfr{al} $d$ is a non-running scale distance and $b$ is a
constant. The best fitted values which reproduce the magnitude of
the anomalous Pioneer acceleration are (Brownstein and Moffat
2006)
\begin{equation}\left\{\begin{array}{lll}
\zeta_{\infty}=(1.00\pm 0.02)\times 10^{-3},\\\\
\lambda_{\infty}=47\pm 1\ {\rm AU},\\\\
d=4.6\pm 0.2 \ {\rm AU},\\\\
b=4.0.
 \lb{fit}\end{array}\right.\end{equation}
The `renormalized' value $G_{\infty}$ of the Newtonian
gravitational constant-$G$ in the following-which is measured by
the usual astronomical techniques is related to the `bare'
constant by (Brownstein and Moffat 2006) \eqi
\rp{G_0}{G_{\infty}}=\rp{1}{ 1+\sqrt{\zeta_{\infty}} }.\eqf With
the fit of \rfr{fit} we have \eqi
\rp{G_0}{G_{\infty}}=0.96934.\eqf

The scope  of Brownstein and Moffat (2006) is to correctly
reproduce the Pioneer anomalous acceleration without contradicting
either the equivalence principle or our knowledge of the planetary
orbital motions. The first requirement is satisfied by the metric
character of their theory. In regard to the second point,
Brownstein and Moffat (2006) do not limit
  the validity of \rfr{accel} just to the region in which the Pioneer
anomaly manifested itself, but extend it to the entire Solar
System. Their model is not a mere more or less $ad\ hoc$ scheme
just to save the phenomena being, instead, rather `rigid' and
predictive. It is an important feature because it, thus, allows
for other tests independent of the Pioneer anomaly itself. This
general characteristic will also be  preserved in future if and
when more points to be fitted will be obtained by further and
extensive re-analysis of the entire data set of the Pioneer
spacecraft (Turyshev et al. 2006a; 2006b) yielding a modification
of the fit of \rfr{fit}. Brownstein and Moffat (2006) perform a
test based on the observable
\eqi\eta=\left[\rp{G(a)}{G(a_{\oplus})}\right]^{1/3}-1,\eqf where
$a$ and $a_{\oplus}$ are the semimajor axes of a planet and the
Earth. The quantity $\eta$ is related to the third Kepler's law
for which observational constraints exist from a previous
model-independent analysis (Talmadge et al. 1988) for the inner
planets and Jupiter. No observational limits were put beyond
Saturn because of the inaccuracy of the optical data used at the
time of the analysis by Talmadge et al. (1988). Brownstein and
Moffat (2006) find their predictions for $\eta$ in agreement with
the data of Talmadge et al. (1988).

As an independent test of the validity of the proposed mechanism,
we will follow an approach similar to that of (Iorio and Giudice
2006) by suitably analyzing the orbital motion of the outer
planets of the Solar System in the context of the latest
observationally determinations by the Russian astronomer E.V.
Pitjeva (Institute of Applied Astronomy, Russian Academy of
Sciences). She has recently processed almost one century of data
of all types in the effort of continuously improving the EPM2004
planetary ephemerides (Pitjeva 2005a). Among other things, she
also determined anomalous secular, i.e. averaged over one orbital
revolution, rates of the perihelia $\Delta\dot\varpi_{\rm deter}$
of the inner (Pitjeva 2005b) and of some of the outer (Pitjeva
2006a; 2006b) planets as fit-for parameters\footnote{The
perihelia, as the other Keplerian orbital elements, are not
directly observable quantities, contrary to, e.g., $\alpha$ and
$\delta$.} of global solutions in which she contrasted, in a
least-square way, the observations (ranges, range-rates, angles
like right ascension $\alpha$ and declination $\delta$, etc.) to
their predicted values computed with a complete suite of dynamical
force models including all the known Newtonian and Einsteinian
features of motion. Thus, any unmodelled force, as it would be the
case for a Pioneer-like one if present in Nature, is entirely
accounted for by the determined perihelia extra-rates. In regard
to the outer planets, Pitjeva was able to determine the
extra-precessions of perihelia for Jupiter, Saturn and Uranus (see
Table \ref{tavolapar} for their relevant orbital parameters)
{\small\begin{table}\caption{ Semimajor axes $a$, in AU,
eccentricities $e$ and orbital periods, in years, of Jupiter,
Saturn and Uranus. For such planets modern data sets covering at
least one orbital revolution exist; for Neptune and Pluto it is
not so. }\label{tavolapar}

\begin{tabular}{llll} \noalign{\hrule height 1.5pt}

 & Jupiter & Saturn  & Uranus\\
$a$ & 5.2 & 9.5 & 19.19\\
$e$ & 0.048 & 0.056 & 0.047\\
$P$ & 11.86 & 29.45 & 84.07\\
\hline

\noalign{\hrule height 1.5pt}
\end{tabular}

\end{table}}
because the temporal extension of the used data set covered at
least one full orbital revolution just for such planets: indeed,
the orbital periods of Neptune and Pluto amount to about 164  and
248 years, respectively. For the external regions of the Solar
System only optical observations have been used (Pitjeva 2005a);
they are, undoubtedly, of poorer accuracy with respect to those
used for the inner planets which also benefit of radar-ranging
measurements, but we will show that they are accurate enough for
our purposes.
\section{Comparison with the observational determinations}
\begin{figure}
\begin{center}
\includegraphics[width=14cm,height=11cm]{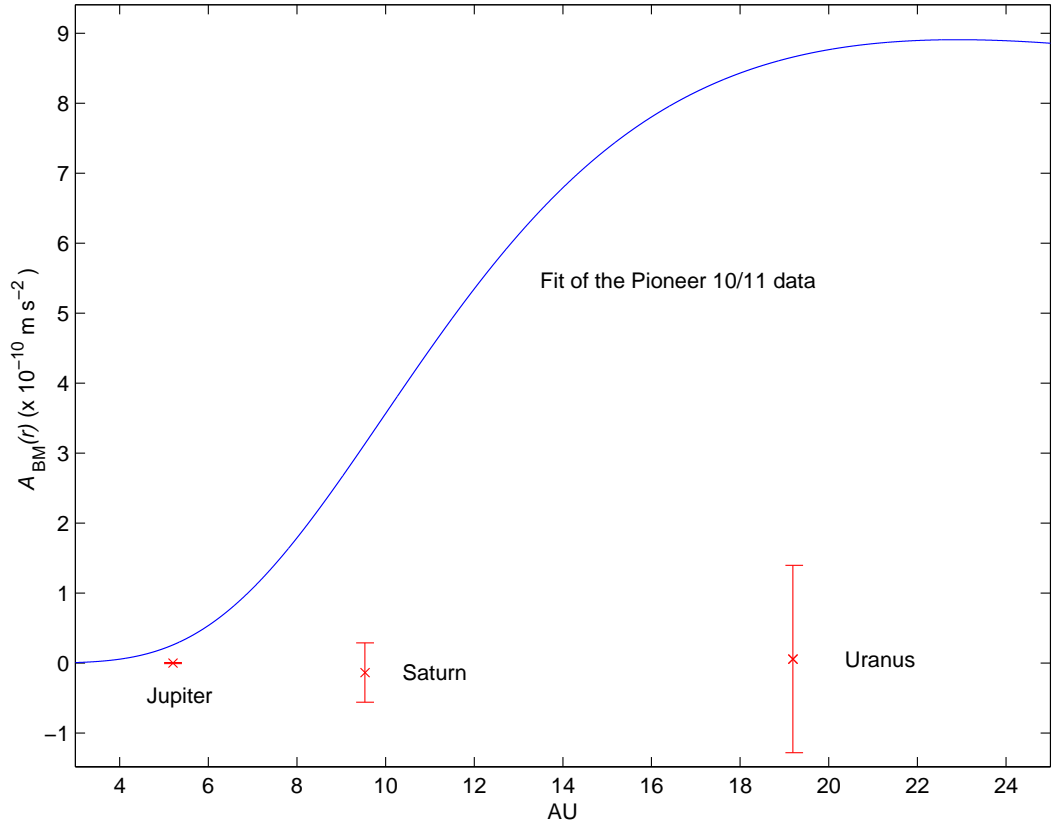}
\end{center}
\caption{\label{figura} The continuous curve is the fit to the
currently available Pioneer 10/11 data according to the model of
\rfr{accel} by Brownstein and Moffat (2006). The determined
anomalous accelerations experienced by Jupiter, Saturn and Uranus,
from the perihelion rates by Pitjeva (2006a; 2006b), are also
shown.}
\end{figure}

In, e.g., (Iorio and Giudice 2006; Sanders 2006) it has been shown
that a radial and constant perturbing acceleration $A$ induces a
pericentre rate \eqi \dert\varpi
t=A\sqrt{\rp{a(1-e^2)}{GM}}.\lb{peri}\eqf We will use \rfr{peri}
and the determined extra-rates of perihelion (Pitjeva 2006a;
2006b) in order to solve for $A$ and compare the so-obtained
values with those predicted by \rfr{accel} for Jupiter, Saturn and
Uranus. The results are summarized in Table \ref{tavola} and
Figure \ref{figura}
{\small\begin{table}\caption{ First row: determined
extra-precessions of the longitudes of perihelia of Jupiter,
Saturn and Uranus, in arcseconds per century (Pitjeva 2006a;
2006b). The quoted uncertainties are the formal, statistical
errors re-scaled by a factor 10 in order to get realistic
estimates. Second row: predicted anomalous acceleration for
Jupiter, Saturn and Uranus, in units of $10^{-10}$ m s$^{-2}$,
according to the model of \rfr{accel} (Brownstein and Moffat
2006), evaluated at $r=a$. Third row: determined anomalous
acceleration of Jupiter, Saturn and Uranus, in units of $10^{-10}$
m s$^{-2}$, from the figures of the first row. The quoted
uncertainties have been obtained by means of the re-scaled errors
in the perihelia rates. Fourth row: discrepancy between the
determined and predicted accelerations in units of errors
$\sigma$. }\label{tavola}

\begin{tabular}{llll} \noalign{\hrule height 1.5pt}

 & Jupiter & Saturn  & Uranus\\
$\Delta\dot\varpi_{\rm deter}$ & $0.0062\pm 0.036$ & $-0.92\pm 2.9$ & $0.57\pm 13.0$\\
$A_{\rm BM}(a)$ & 0.260 & 3.136 & 8.660\\
$A_{\rm deter}$  & $0.001\pm 0.007$ & $-0.134\pm 0.423$ & $0.058\pm 1.338$ \\
$|A_{\rm deter}-A_{\rm BM}(a)|/\sigma$& 37 & 7 &6\\

\hline

\noalign{\hrule height 1.5pt}
\end{tabular}

\end{table}}
As can be noted, the model proposed by Brownstein and Moffat
(2006), in the form of \rfr{accel} and with the fitted values of
\rfr{fit}, must be rejected. Note that the quoted errors for the
perihelia rates are the formal uncertainties multiplied by 10 in
order to give conservative evaluations of the realistic ones. In
the case of Jupiter even a re-scaling of 100 would still reject
the value predicted by \rfr{accel}.

More generally, let us now forget any particular gravitational
model which may be able to accommodate the Pioneer anomaly and
assume, in a pure phenomenological way as done as in (Iorio and
Giudice 2006), that an anomalous acceleration like $A_{\rm Pio}$
of \rfr{pioa} acts upon the planets in the outer regions of the
Solar System;  \rfr{peri} and \rfr{pioa} yield an anomalous rate
of $83.58\pm 12.71$ arcseconds per century for Uranus. Table
\ref{tavola} tells us that such a prediction would be ruled out by
the determined extra-rates of perihelia even by multiplying the
formal error for Uranus (1.3 arcseconds per century) by 50. This
result is  very important because it is unlikely that a further,
extensive re-analysis of the entire Pioneer 10/11 data set
(Turyshev et al. 2006a; 2006b) will substantially change our
knowledge of the Pioneer anomaly in the regions crossed by Uranus,
being focussed on shedding light on what occurred to both the
spacecraft well below 20 AU.

The results presented here are consistent with the findings of
(Iorio and Giudice 2006) in which the time-dependent patterns of
the true observable quantities  like $\alpha\cos\delta$ and
$\delta$ induced by a Pioneer-like acceleration on Uranus, Neptune
and Pluto have been compared with the observational residuals
determined in (Pitjeva 2005a) for the same quantities and the same
planets over a time span of about 90 years from 1913 (1914 for
Pluto) to 2003. While the former ones exhibited well defined
polynomial signatures yielding shifts of hundreds of arcseconds,
the latter ones did not show any particular patterns, being almost
uniform strips constrained well within $\pm 5$ arcseconds over the
data set time span which includes the entire Pioneer 10/11
lifetimes. An analogous conclusion can also be found in (Tangen
2006), although a different theoretical quantity has been used in
the comparison with the data.

\section{Conclusions}
In this paper we have used the latest observational determinations
of the perihelion rates of Jupiter, Saturn and Uranus by E.V.
Pitjeva (2006a; 2006b) to perform an unambiguous and independent
test of the gravitational mechanism for explaining the Pioneer
anomaly recently proposed by Brownstein and Moffat (2006) on the
basis of the scalar-tensor-vector (STVG) metric theory of gravity
by Moffat (2006; 2007). It turns out that the values predicted by
the STVG model, fitted to all the currently available Pioneer
10/11 data, for the anomalous accelerations experienced by
Jupiter, Saturn and Uranus are neatly contradicted by those
obtained from the determined extra-rates of perihelia for the same
planets, even by conservatively accounting for the fact that the
used observations for them are only optical and of modest
precision with respect to those for the inner planets. The result
for Uranus, also confirmed by a purely phenomenologically,
model-independent analysis, is very important because it seems
unlikely that the planned further and extensive re-analysis of the
entire Pioneer 10/11 data set (Turyshev et al. 2006a; 2006b) will
modify what we already know about the Pioneer anomaly in the
region  $\gtrsim 20$ AU in a relevant manner for our conclusions.

Thus, the hypothesis that the anomalous behavior of the Pioneer
spacecraft can find some explanation of gravitational origin
further weaken.

\section*{Acknowledgments}
I am grateful to E.V. Pitjeva (Institute of Applied Astronomy,
Russian Academy of Sciences) for her results about the outer
planets of the Solar System and related discussions.

\end{document}